\documentclass[english,prd,preprint,nofootinbib]{revtex4}
\usepackage{times}
\usepackage[T1]{fontenc}
\usepackage[latin1]{inputenc}
\usepackage{amsmath}
\usepackage{graphicx}
\usepackage{amssymb}

\makeatletter

\usepackage{color}

\makeatletter

\usepackage{color}



\makeatother

\usepackage{babel}
\makeatother
\begin{document}
\renewcommand{\thefootnote}{\#\arabic{footnote}} \newcommand{\rem}[1]{{\bf [#1]}}
\newcommand{\gsim}{ \mathop{}_ {\textstyle \sim}^{\textstyle >} }
\newcommand{\lsim}{ \mathop{}_ {\textstyle \sim}^{\textstyle <} }
\newcommand{\vev}[1]{ \left\langle {#1}  \right\rangle } \newcommand{\bear}{\begin{array}}
\newcommand{\eear}{\end{array}} \newcommand{\bea}{\begin{eqnarray}}
\newcommand{\eea}{\end{eqnarray}} \newcommand{\beq}{\begin{equation}}
\newcommand{\eeq}{\end{equation}} \newcommand{\bef}{\begin{figure}}
\newcommand{\eef}{\end{figure}} \newcommand{\bec}{\begin{center}}
\newcommand{\eec}{\end{center}} \newcommand{\non}{\nonumber} \newcommand{\eqn}[1]{\beq {#1}\eeq}
\newcommand{\la}{\left\langle} \newcommand{\ra}{\right\rangle} \newcommand{\ds}{\displaystyle}
\def\SEC#1{Sec.~\ref{#1}} \def\FIG#1{Fig.~\ref{#1}} \def\EQ#1{Eq.~(\ref{#1})}
\def\EQS#1{Eqs.~(\ref{#1})} \def\lrf#1#2{ \left(\frac{#1}{#2}\right)}
\def\lrfp#1#2#3{ \left(\frac{#1}{#2} 
\right)^{#3}} \def\GEV#1{10^{#1}{\rm\,GeV}} \def\MEV#1{10^{#1}{\rm\,MeV}} \def\KEV#1{10^{#1}{\rm\,keV}}

\def\ap{A^\prime}


\begin{flushright}IPMU 08-0058 \par\end{flushright}

\title{Gamma rays and positrons from a decaying hidden gauge boson}

\author{Chuan-Ren Chen$^{1}$, Fuminobu Takahashi$^{1}$ and T. T. Yanagida$^{1,2}$}

\affiliation{${}^{1}$Institute for the Physics and Mathematics of the Universe,
University of Tokyo, Chiba 277-8568, Japan\\
 ${}^{2}$Department of Physics, University of Tokyo, Tokyo 113-0033,
Japan }

\begin{abstract}
We study a scenario that a hidden gauge boson constitutes the dominant
component of dark matter and decays into the standard model particles
through a gauge kinetic mixing. Interestingly, gamma rays and positrons
produced from the decay of hidden gauge boson can explain both the
EGRET excess of diffuse gamma rays and the HEAT anomaly in the positron
fraction. The spectra of the gamma rays and the positrons have distinctive
features; the absence of line emission of the gamma ray and a sharp
peak in the positron fraction. Such features may be observed by the
FGST and PAMELA satellites. 
\end{abstract}
\maketitle

\section{introduction}

\label{sec:1} 
The concept of symmetry has been the guiding principle in modern physics.
The structure of the standard model (SM) is dictated by $SU(3)_{C}\times SU(2)_{L}\times U(1)_{Y}$
gauge symmetries. The electroweak symmetry, $SU(2)_{L}\times U(1)_{Y}$,
is spontaneously broken by a non-vanishing vacuum expectation value
(vev) of the Higgs boson, and massive $W$ and $Z$ bosons are generated.
It is quite natural in the string landscape that there are many other
gauge symmetries as well as discrete ones realized in nature, and
some of the gauge symmetries may be spontaneously broken, leading
to massive gauge bosons, as in the SM.

Suppose that some of the hidden gauge bosons are `odd' under a $Z_{2}$
parity, whereas all the SM particles are `even'. If the parity is
exact, and if the hidden gauge symmetries are spontaneously broken,
the lightest parity-odd gauge boson is stable and can be a good candidate
for cold dark matter. Some recent examples are the T-odd $U(1)$ gauge
boson in the little Higgs model with T-parity~\cite{ArkaniHamed:2002qy,Cheng:2003ju,Cheng:2004yc,Csaki:2008se}
and the KK photon in the minimal universal extra dimension model~\cite{Appelquist:2000nn}.
However, it is not known yet which of unbroken or (explicitly or spontaneously)
broken discrete symmetries are more common in the string landscape.
If a discrete symmetry breaking is a general phenomenon, we may expect
that the dark matter is not absolutely stable and ultimately decays
into the SM particles.

In this paper we consider a simplest case that there exist a hidden
$U(1)^{\prime}$ gauge symmetry which is spontaneously broken and
a parity under which only the hidden gauge boson changes its sign.
We assume that the parity is broken by a kinetic mixing between the
$U(1)^{\prime}$ and the SM $U(1)_{Y}$ gauge symmetries\ \cite{Holdom:1985ag,Foot:1991kb}.
As a result, the hidden gauge boson decays into the SM particles through
the parity violating interactions induced by the kinetic mixing. If
such a violation is so tiny that the lifetime of the hidden gauge
boson is much longer than the age of our universe, the hidden gauge
boson can be the dominant component of dark matter. Furthermore, the
subsequent decays of the SM particles will form continuous spectra
of the gamma rays and the positrons in the high-energy cosmic ray.
With an appropriate amount of the parity violation, we see that those
gamma rays and positrons could be the source of the excesses of the
gamma ray observed by Energetic Gamma Ray Experiment Telescope (EGRET)~\cite{Sreekumar:1997un,Strong:2004ry}
and the positron flux observed by High Energy Antimatter Telescope
(HEAT)\ \cite{Barwick:1997ig}, MASS\ \cite{Grimani:2002yz} and
AMS\ \cite{Aguilar:2007yf} experiments.

Recently, the gravitino dark matter scenario was extensively studied
in the framework of supersymmetry with $R$-parity violation~\cite{Takayama:2000uz,Buchmuller:2007ui}.
The decay of the gravitino into the gamma rays was discussed in Refs.~\cite{Takayama:2000uz,Buchmuller:2007ui}.
Furthermore, if the gravitino is heavier than $W$ boson, it decays
predominantly into a $W$ or $Z$ boson and a lepton. With the gravitino
mass of ${\cal O}(100)$ GeV and the lifetime of ${\cal O}(10^{26})$
sec, the gravitino decay can simultaneously explain the EGRET anomaly
in the extragalactic diffuse gamma ray background and the HEAT excess
in the positron fraction\ \cite{Bertone:2007aw,Ibarra:2007wg,Ibarra:2008qg,Ishiwata:2008cu}.
Our decaying hidden gauge boson has therefore many parallels with
the gravitino with $R$-parity violation. However, in our model, the
decay branching ratio of the hidden gauge boson to $W$ boson is highly
suppressed, which is significantly different from the gravitino case.

This paper is organized as follows. In Sec.\ \ref{sec:model}, we
present the effective Lagrangian and Feynman rules used in our calculations.
In Sec.\ \ref{sec:numerical}, we calculate the spectra of gamma
ray and positron flux from the decay of the hidden gauge boson and
compare them with the observed data. Sec.\ \ref{sec:Conclusion}
is devoted to discussions and conclusions.


\section{Framework}

\label{sec:model} 
We consider a hidden Abelian gauge symmetry $U(1)^{\prime}$ and the
associated gauge boson $B_{\mu}^{\prime}$, and introduce a hidden
parity under which $B_{\mu}^{\prime}$ transforms as\begin{equation}
B_{\mu}^{'}\to-B_{\mu}^{'}.\,\label{eq:parity}\end{equation}
 The $U(1)^{\prime}$ symmetry is assumed to be spontaneously broken,
so that the gauge boson $B_{\mu}^{\prime}$ has a non-vanishing mass,
$m$.

We assume that the low energy effective theory can be written in terms
of the SM particles and the hidden gauge boson $B_{\mu}^{\prime}$,
and that all the SM particles are neutral under both the $U(1)^{\prime}$
and the hidden parity. We would like to introduce a tiny parity violating
interaction. Among many possibilities, we will focus on a kinetic
mixing term between $U(1)^{\prime}$ and $U(1)_{Y}$, since the kinetic
mixing has the lowest dimension~%
\footnote{It is also possible to write down non-renormalizable parity violating
interactions suppressed by a large mass scale. For the magnitudes
of kinetic mixing in our work, however, we can safely neglect such
interactions if they are suppressed by the Planck scale. %
}. The low energy effective Lagrangian can be written as 
\begin{equation}
{\cal L}\;\supset\;-\frac{1}{4}B_{\mu\nu}B^{\mu\nu}-\frac{1}{4}B_{\mu\nu}^{\prime}B^{\prime\mu\nu}+\frac{\epsilon}{2}B_{\mu\nu}B^{\prime\mu\nu}+\frac{1}{2}m^{2}B_{\mu}^{\prime}B^{\prime\mu},\,\label{eq:kinetic}\end{equation}
where $\epsilon(\ll1)$ is the coefficient of the hidden parity violation
term, and $B_{\mu}$ is the gauge boson of $U(1)_{Y}$, $B_{\mu\nu}$
and $B_{\mu\nu}^{\prime}$ are the field strength of $U(1)_{Y}$ and
$U(1)^{\prime}$, respectively: $B_{\mu\nu}^{(\prime)}=\partial_{\mu}B_{\nu}^{(\prime)}-\partial_{\nu}B_{\mu}^{(\prime)}$.
In the following discussion we neglect ${\cal O}(\epsilon^{2})$ terms,
since $\epsilon$ must be extremely small as ${\cal O}(10^{-26})$
for the hidden gauge boson to become dark matter. We will also give
a possible origin of such tiny kinetic mixing later.

We can remove the kinetic mixing and bring the kinetic terms into
canonical form by redefining the gauge fields as~\cite{Fukuda:1974rd}
\begin{eqnarray}
\tilde{B} & = & B-\epsilon B^{'},\nonumber \\
\tilde{B}' & = & B'\,.\,\label{eq:redifine}\end{eqnarray}
 Note that we omit the Lorentz index $\mu$ hereafter. Taking account
of the electroweak symmetry breaking, the mass terms of neutral gauge
bosons are 
\begin{equation}
{\cal L}_{mass}^{Gauge}\;=\;\frac{v^{2}}{8}\,(W^{3},\tilde{B},\tilde{B}^{\prime})\left(\begin{array}{ccc}
g^{2}, & -gg^{\prime}, & -gg^{\prime}\epsilon\\
-gg^{\prime}, & g^{\prime2}, & g^{\prime2}\epsilon\\
-gg^{\prime}\epsilon, & g^{\prime2}\epsilon, & 4m^{2}/v^{2}\end{array}\right)\left(\begin{array}{c}
W^{3}\\
\tilde{B}\\
\tilde{B}^{\prime}\end{array}\right),\label{eq:mass_mtx}\end{equation}
where $W^{3}$ is the neutral gauge field of $SU(2)_{L}$, $v$ is
the vev of the doublet Higgs field $H$, $g$ and $g^{\prime}$ are
weak and hypercharge coupling constants, respectively. After diagonalizing
the mass matrix, we can express the gauge eigenstates $W^{3}$, $B$
and $B^{\prime}$ in terms of the mass eigenstates $Z$, $A$, and
$A^{\prime}$ as 
\begin{align}
W^{3} & =c_{W}Z+s_{W}A-c_{W}s_{W}\frac{m_{Z}^{2}}{m_{A'}^{2}-m_{Z}^{2}}\epsilon A^{\prime},\nonumber \\
B & =-s_{W}Z+c_{W}A+\frac{m_{A'}^{2}-c_{W}^{2}m_{Z}^{2}}{m_{A'}^{2}-m_{Z}^{2}}\epsilon A^{\prime},\,\label{eq:diagonal}\\
B^{\prime} & =s_{W}\frac{m_{Z}^{2}}{m_{A'}^{2}-m_{Z}^{2}}\epsilon Z+A^{\prime}\nonumber \end{align}
with 
\begin{align}
c_{W} & \equiv\cos\theta_{W}=g/\sqrt{g^{2}+g^{\prime2}},\quad s_{W}\equiv\sin\theta_{W}=\sqrt{1-c_{W}^{2}},\nonumber \\
m_{Z}^{2} & =\frac{1}{4}(g^{2}+g^{\prime2})v^{2},\quad m_{A^{\prime}}^{2}=m^{2},\,\label{eq:mass}\end{align}
where $m_{A^{\prime}}$ and $m_{Z}$ are the masses of $A^{\prime}$
and $Z$, respectively. One can easily see that $A$ and $Z$ are
reduced to the ordinary photon and $Z$ boson in the limit of $\epsilon\rightarrow0$.
Since interested values of $\epsilon$ are extremely small, the kinetic
mixing hardly affects any SM predictions of the electroweak measurements.

The $A^{\prime}$ interacts with the SM particles through the mixings
shown in Eq.~(\ref{eq:diagonal}). The Feynman rules can be derived
in a straightforward way by expanding the SM Lagrangian with respect
to $\epsilon$\ \cite{Chang:2006fp,Gopalakrishna:2008dv}, and the
results are 
\begin{align}
\bar{u}A_{\mu}^{\prime}u & :-i\frac{g^{\prime}\epsilon}{2(m_{A'}^{2}-m_{Z}^{2})}\gamma_{\mu}\left(-\frac{4}{3}c_{W}^{2}m_{Z}^{2}+\frac{5}{6}m_{A'}^{2}+\frac{1}{2}m_{A'}^{2}\gamma_{5}\right),\nonumber \\
\bar{d}A_{\mu}^{\prime}d & :-i\frac{g^{\prime}\epsilon}{2(m_{A'}^{2}-m_{Z}^{2})}\gamma_{\mu}\left(\frac{2}{3}c_{W}^{2}m_{Z}^{2}-\frac{1}{6}m_{A'}^{2}-\frac{1}{2}m_{A'}^{2}\gamma_{5}\right),\nonumber \\
\bar{\nu}A_{\mu}^{\prime}\nu & :i\frac{g^{\prime}\epsilon m_{A'}^{2}}{4(m_{A'}^{2}-m_{Z}^{2})}\gamma_{\mu}(1-\gamma_{5}),\nonumber \\
\bar{e}A_{\mu}^{\prime}e & :-i\frac{g^{\prime}\epsilon}{4(m_{A'}^{2}-m_{Z}^{2})}\gamma_{\mu}\left(4c_{W}^{2}m_{Z}^{2}-3m_{A'}^{2}-m_{A'}^{2}\gamma_{5}\right),\nonumber \\
WWA^{\prime} & :-\frac{s_{W}m_{Z}^{2}\epsilon}{m_{A'}^{2}-m_{Z}^{2}}g_{WWZ}^{SM},\label{eq:couplings}\end{align}
where $u$, $d$, $e$ and $\nu$ represent all the up-type quarks,
down-type quarks, charged leptons and neutrinos, respectively, and
$g_{WWZ}^{SM}$ is the coupling of $WWZ$ in the SM.

\begin{figure}
\includegraphics[scale=0.5]{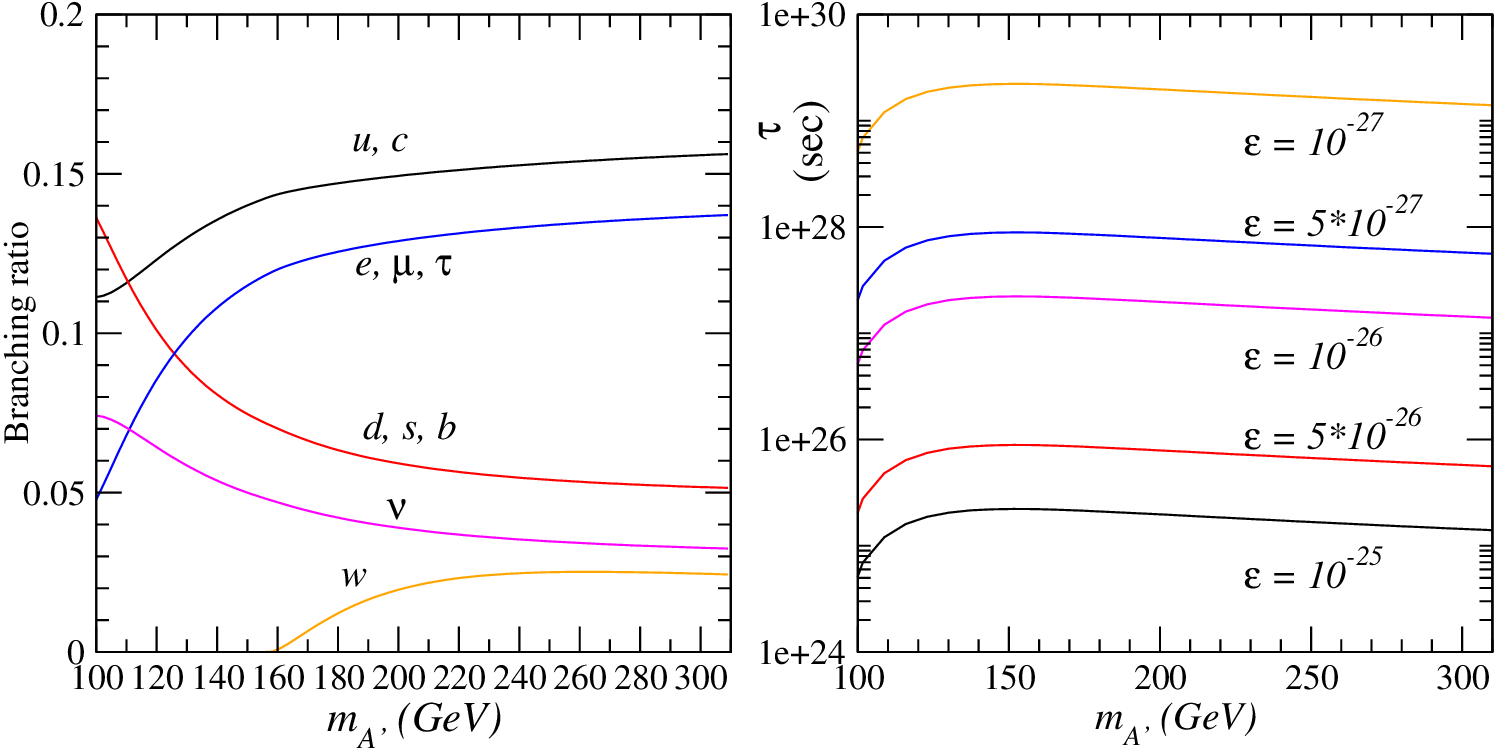}

\caption{Decay branching ratios and lifetime of $A^{\prime}$, as a function
of $m_{A'}$.\ \label{fig:decay}}
\end{figure}

The hidden gauge boson $\ap$ decays into the SM particles through
the above interactions. Fig.~\ref{fig:decay} shows the decay branching
ratios and the lifetime of $A^{\prime}$ as a function of $m_{A'}$.
The down-type quark decay modes of $A^{\prime}$ , $A'\to d\bar{d}\,(d=d,s,b)$,
dominate over the other modes for $m_{A'}\simeq100$ GeV, whereas
they decrease quickly as $m_{A'}$ increases. For $m_{A'}\gtrsim120$
GeV, the up-type quark decay modes, $A'\to u\bar{u}\,(u=u,c)$, become
the largest ones followed by charged lepton decay modes, $A'\to\ell^{+}\ell^{-}\,(\ell=e,\mu,\tau)$.
This behavior can be understood easily as follows. The $A^{\prime}$
becomes more like the $Z$ boson as $m_{A'}$ approaches $m_{Z}$,
and therefore, the partial decay width of $A^{\prime}\rightarrow f\bar{f}$
is proportional to $g_{V}^{2}+g_{A}^{2}$, where $g_{V(A)}$ is the
vector (axial) coupling strength of the neutral weak interaction.
On the other hand, as $m_{A'}$ becomes much heavier than $m_{Z}$,
$A^{\prime}$ tends to not feel the electroweak symmetry breaking,
so the branching ratios are insensitive to $m_{A'}$. For \textcolor{red}{}$m_{\ap}\gtrsim350{\rm \, GeV}$,
the decay mode of $A'\to t\bar{t}$ is allowed. The following discussions
on the gamma rays and the positrons are not significantly modified
even for $m_{\ap}\gtrsim350{\rm \, GeV}$.

\section{Gamma ray and positron spectra}

\label{sec:numerical} 
The hidden gauge boson $A^{\prime}$ decays into a pair of SM fermions
and $W$ bosons if kinematically allowed, as we have seen in the previous
section. Due to the subsequent QCD hadronization processes, a bunch
of hadrons are produced, and in particular, continuum spectra for
the photon and the positron are formed. Throughout this paper we assume
that the $A^{\prime}$ constitutes the dominant component of dark
matter. Then the produced photons and positrons may be observed in
the high-energy cosmic rays. In this section we estimate the fluxes
of the gamma rays and the positrons produced from the $A'$ decay,
and see how they may account for the observed excesses.


\subsection{Gamma-ray flux}

\begin{figure}
\includegraphics[scale=0.4]{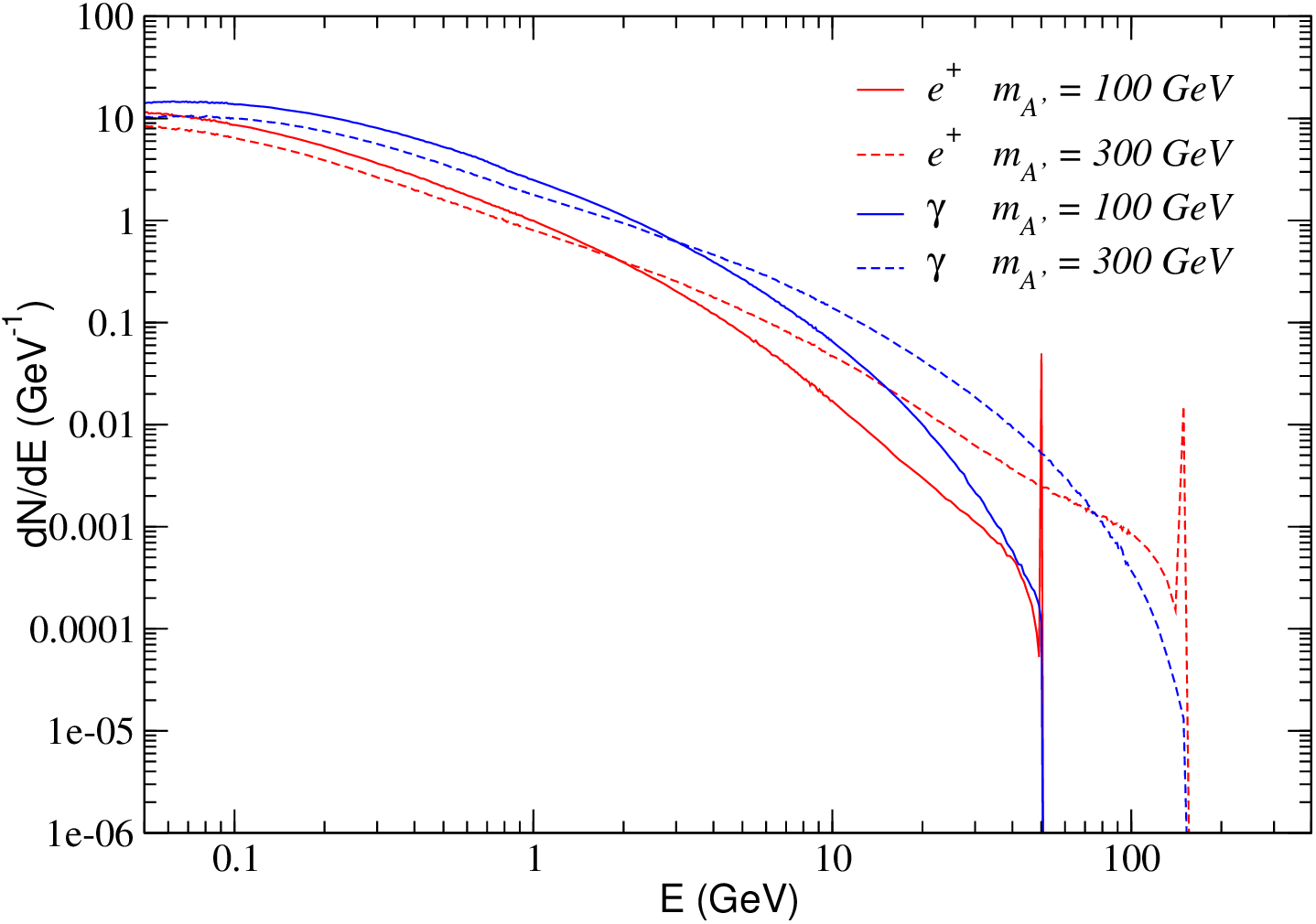}

\caption{Energy spectra of $\gamma$ and $e^{+}$ generated from the decay
of $A'$.~\label{fig:dnde}}
\end{figure}

The gamma-ray energy spectrum is characterized by $dN_{\gamma}/dE$,
the number of photons having energy between $E$ and $E+dE$, produced
from the decay of one $A^{\prime}$ gauge boson. The main contribution
to the continuous spectrum of $\gamma$ arises from the $\pi^{0}$
generated in the QCD hadronization process. To estimate the spectrum,
we use the PYTHIA\ \cite{Sjostrand:2006za} Monte Carlo program with
the branching ratios shown in Fig.~\ref{fig:decay}. We also include
the real gamma-ray emission, known as internal bresstrahlung, from
the charged particles from the $A'$ direct decay~%
\footnote{We are grateful to John Beacom\ \cite{Beacom:2004pe} for bringing
up this fact to us.%
}, whose contributions to $dN_{\gamma}/dE$ become important compared
to that from $\pi^{0}$ when $E\to m_{A'}/2$. However, the features
of the gamma-ray flux which we discuss below will not change even
though the internal bresstrahlung effects are not included. The energy
spectra $dN_{\gamma}/dE$ for $m_{\ap}=100$ and $300$ GeV are shown
in Fig.~\ref{fig:dnde}. It is worth noting that there is no line
emission of the gamma rays from the decay of $A^{\prime}$, which
is present in the case of the gravitino dark matter.

There are galactic and extragalactic contributions from the decay
of $\ap$ to the observed gamma ray flux. The flux of the gamma ray
from the extragalactic origin is estimated as\ \cite{Ibarra:2007wg,Ishiwata:2008cu}
\begin{equation}
\left[E^{2}\frac{dJ_{\gamma}}{dE}\right]_{eg}=\frac{E^{2}c\Omega_{A'}\rho_{c}}{4\pi m_{A'}\tau_{A'}H_{0}\Omega_{M}^{1/2}}\int_{1}^{y_{{\rm eq}}}dy\frac{dN_{\gamma}}{d(yE)}\frac{y^{-3/2}}{\sqrt{1+\frac{\Omega_{\Lambda}}{\Omega_{M}}y^{-3}}},\label{eq:gamma_eg}\end{equation}
where $c$ is the speed of light; $\Omega_{A'}$, $\Omega_{M}$ and
$\Omega_{\Lambda}$ are the density parameters of $A'$, matter (including
both baryons and dark matter) and the cosmological constant, respectively;
$\rho_{c}$ is the critical density; $\tau_{A'}$ is the lifetime
of $A'$; $H_{0}$ is the Hubble parameter at the present time; $y\equiv1+z$,
where $z$ is the redshift, and $y_{{\rm eq}}$ denotes a value of
$y$ at the matter-radiation equality. For the numerical results,
we use\ \cite{Komatsu:2008hk}\begin{equation}
\Omega_{A'}h^{2}=0.1099,\quad\Omega_{M}h^{2}=0.1326,\quad\Omega_{\Lambda}=0.742,\quad\rho_{c}=1.0537\times10^{-5}{\rm GeV/cm}^{3}.\,\label{eq:input}\end{equation}

On the other hand, the gamma ray flux from the decay of $\ap$ in
the Milky Way halo is 
\begin{equation}
\left[E^{2}\frac{dJ_{\gamma}}{dE}\right]_{halo}=\frac{E^{2}}{4\pi m_{A'}\tau_{A'}}\frac{dN_{\gamma}}{dE}\left\langle \int_{los}\rho_{halo}(\vec{\ell})d\vec{\ell}\right\rangle ,\label{eq:gamma_halo}\end{equation}
 where $\rho_{halo}$ is the density profile of dark matter in the
Milky Way, $\left\langle \int_{los}\rho_{halo}(\vec{\ell})d\vec{\ell}\right\rangle $
is the average of the integration along the line of sight (los). We
adopt the Navarro-Frenk-White (NFW) halo profile\ \cite{Navarro:1995iw}
\begin{equation}
\rho(r)=\frac{\rho_{0}}{(r/r_{c})(1+r/r_{c})^{2}}\label{eq:profile}\end{equation}
in our calculation, where $r$ is the distance from the center of
Milky Way, $r_{c}=20$ kpc, and $\rho_{0}$ is set in such a way that
the dark matter density in the solar system satisfies $\rho(r_{\odot})=0.30\:{\rm GeV/cm}^{3}$\ \cite{Bergstrom:1997fj}
with $r_{\odot}=8.5$ kpc being the distance from the Sun to the Galactic
Center.

\begin{figure}
\includegraphics[scale=0.4]{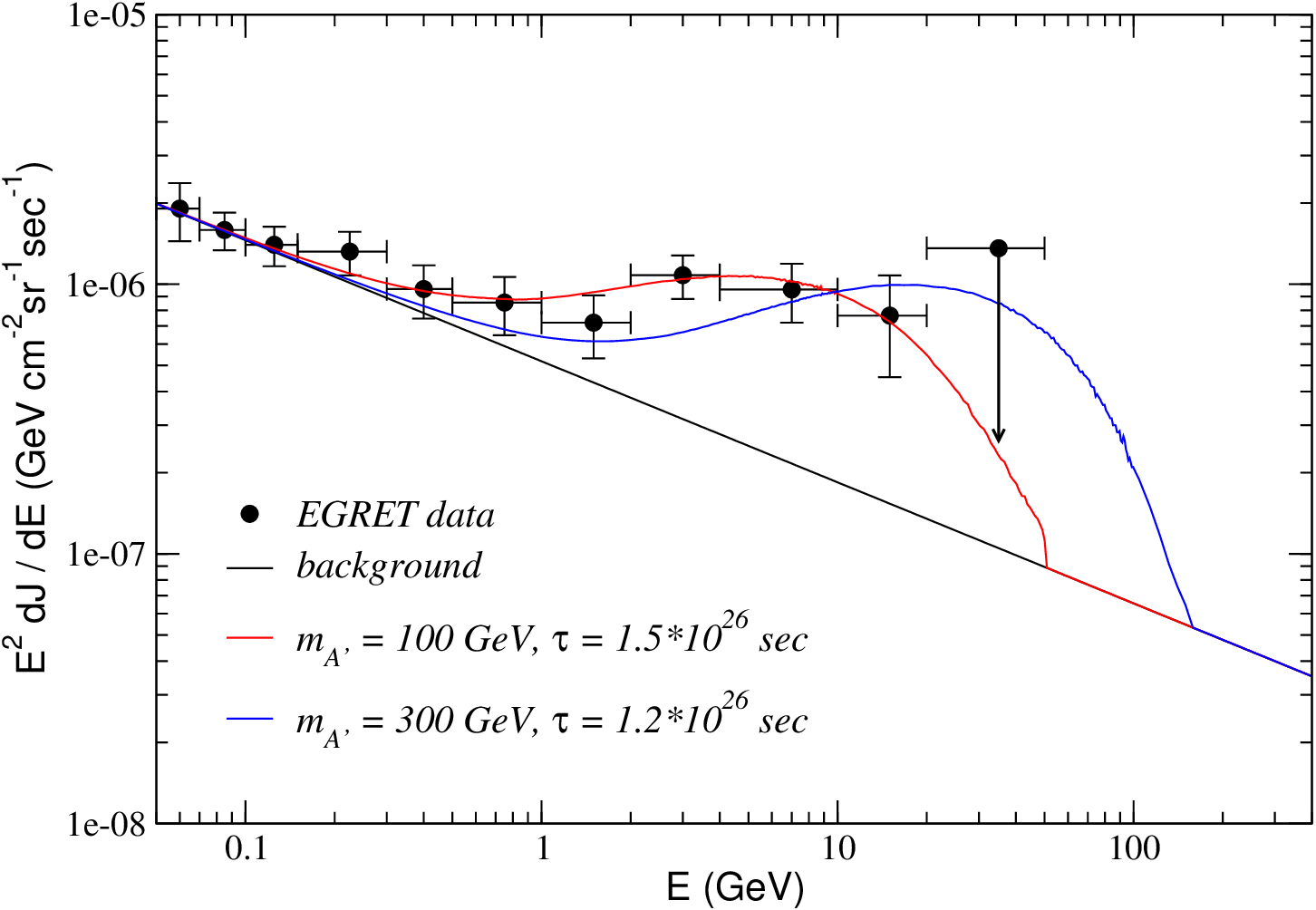}

\caption{$\gamma$ ray flux predicted from decay of $A'$, shown together
with the EGRET data \cite{Strong:2004ry}.\ \label{fig:gamma}}
\end{figure}

In order to compare the EGRET results with the above flux from $A'$
decay, we integrate over the whole sky except for the zone of the
Galactic plane (i.e. the region with the galactic latitudes $|b|<10^{\circ}$).
For the background, we use a power-law form adopted in Ref.\ \cite{Ishiwata:2008cu}
\begin{equation}
\left[E^{2}\frac{dJ_{\gamma}}{dE}\right]_{bg}\simeq5.18\times10^{-7}E^{-0.499}\quad{\rm GeVcm}^{-2}{\rm sr}^{-1}{\rm sec}^{-1},\label{eq:gamma_bg}\end{equation}
 where $E$ is in units of GeV. 
Fig.\ \ref{fig:gamma} shows our numerical results. Recall that there
are only two parameters in our model, i.e. the mass and lifetime of
$A'$, and the peak position only depends on the mass while the deviation
from background is sensitive to both. The signal of \textbf{$m_{A'}=100$}
GeV (the red line) peaks at around $E=5$ GeV region, which is consistent
very well with the observed data. For $m_{A'}=300$ GeV (the blue
line), our prediction can still account for the observed excess, although
the fit is not so good in some energy region. However, given the large
errors in the observed data, it may be premature to extract any sensible
constraint on the model parameters. With more precise data from the
FGST experiment, we should be able to have more information about
the $A'$ mass and lifetime, if the excess is indeed from $A'$ dark
matter decay. As mentioned before, there is no line emission of the
gamma rays in our model, because the production of a pair of on-shell
$\gamma$ from $A'$ is forbidden\ \cite{Yang:1950rg,Hagiwara:1986vm}.
Therefore, there is no secondary peak around the high end of signal
region, which is a characteristic difference between our prediction
and the gravitino case.

\subsection{Positron fraction}

After being produced from the $A'$ decay, the positron will propagate
in the magnetic field of the Milky Way. The typical gyroradius is
much smaller than the size of the galaxy, and the positron will propagate
along the magnetic field. However, since the magnetic fields are tangled,
the motion of the positron can be described by a diffusion equation.
Neglecting the convection and annihilation in the disk, the steady
state solution must satisfy~\cite{Ibarra:2008qg} 
\begin{equation}
\bigtriangledown\cdot\left[K(E,\vec{r})\bigtriangledown f_{e^{+}}\right]+\frac{\partial}{\partial E}\left[b(E,\vec{r})f_{e^{+}}\right]+Q(E,\vec{r})=0,\,\label{eq:e_prop}\end{equation}
where $f_{e^{+}}$ is the number density of $e^{+}$ per unit kinetic
energy, $K(E,\vec{r})$ is the diffusion coefficient, $b(E,\vec{r})$
is the rate of energy loss and $Q(E,\vec{r})$ is the source of producing
$e^{+}$ from $A'$ decay. In our case,\[
Q(E,\vec{r})=\frac{\rho(\vec{r})}{m_{A'}\tau_{A'}}\frac{dN_{e^{+}}}{dE},\]
 where $dN_{e^{+}}/dE$ is the energy spectrum of $e^{+}$ from $A'$
decay obtained by using PYTHIA\ \cite{Sjostrand:2006za} (see Fig.\ \ref{fig:dnde}).
We notice that because of the decay channel of $A'\to e^{+}e^{-}$,
there is a sharp peak at $E=m_{A'}/2$.

\begin{figure}
\includegraphics[scale=0.4]{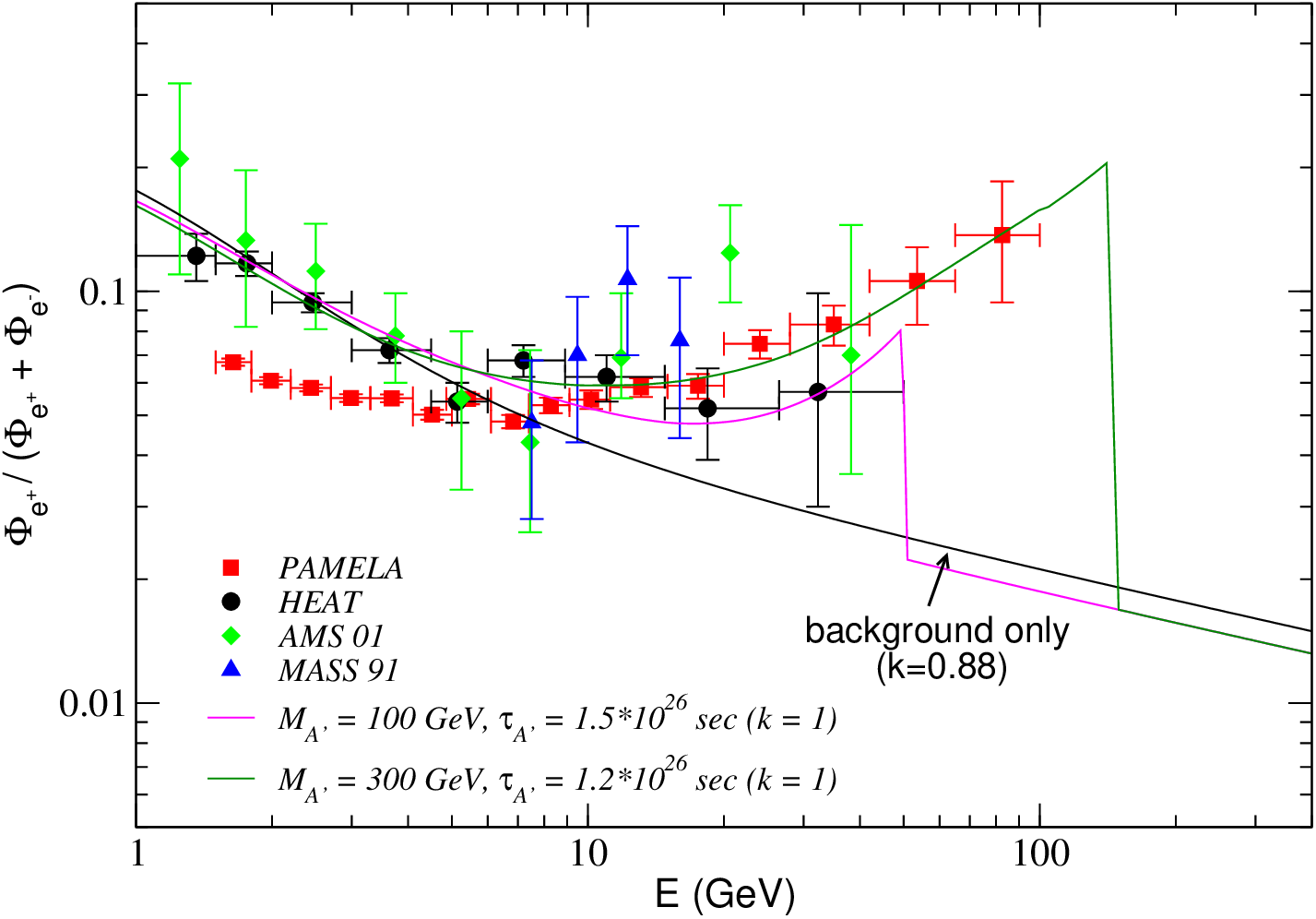}

\caption{Fraction of $e^{+}$ flux from $A'$ decay, shown together with experimental
data\ \cite{Barwick:1997ig,Grimani:2002yz,Aguilar:2007yf,Adriani:2008zr}.\ \label{fig:Fraction_e}}
\end{figure}

The solution of Eq.\ (\ref{eq:e_prop}) in the solar system can be
expressed as 
\begin{equation}
f_{e^{+}}(E)=\frac{1}{m_{A'}\tau_{A'}}\int_{0}^{E_{Max}}dE'G(E,E')\frac{dN_{e^{+}}}{dE'},\label{eq:f_e}\end{equation}
 where $E_{Max}=m_{A'}/2$, and $G(E,E')$ is approximately given
by\ \cite{Ibarra:2008qg} 
 \begin{equation}
G(E,E')\simeq\frac{10^{16}}{E^{2}}e^{a+b(E^{\delta-1}-E'^{\delta-1})}\theta(E'-E)\quad{\rm sec/cm}^{3},\,\label{eq:G_approx}\end{equation}
 where $E$ is in units of GeV, $\delta$ is related to the properties
of the interstellar medium and can be determined mainly from the ratio
of Born to Carbon (B/C)\ \cite{Maurin:2001sj}. We adopt parameters,
$\delta=0.55$, $a=-0.9716$ and $b=-10.012$\ \cite{Ibarra:2008qg},
that are consistent with the B/C value and produce the minimum flux
of positrons. Finally, the flux of $e^{+}$ is given by 
\begin{equation}
\Phi_{e^{+}}^{prim}(E)=\frac{c}{4\pi}f_{e^{+}}=\frac{c}{4\pi m_{A'}\tau_{A'}}\int_{0}^{m_{A'}/2}dE'G(E,E')\frac{dN_{e^{+}}}{dE'}.\,\label{eq:flux_e}\end{equation}
 In addition to $e^{+}$ flux from dark matter decay, there exists
a secondary $e^{+}$ flux from interactions between cosmic rays and
nuclei in the interstellar medium. The positron flux is considered
to be suffered from the solar modulation, especially for the energy
below $10$ GeV. If the solar modulation effect is independent of
the charge-sign, one can cancel the effect by measuring the positron
fraction,\begin{equation}
\frac{\Phi_{e^{+}}}{\Phi_{e^{+}}+\Phi_{e^{-}}}.\,\label{eq:ep_exp}\end{equation}
 Indeed, most of experiments measured the fraction of positron flux.
To estimate the positron fraction, it is necessary to include the
$e^{-}$ flux. We use the approximations of the $e^{-}$ and $e^{+}$
background fluxes \ \cite{Moskalenko:1997gh,Baltz:1998xv} 
\begin{align}
\Phi_{e^{-}}^{prim}(E) & =\frac{0.16E^{-1.1}}{1+11E^{0.9}+3.2E^{2.15}}\quad{\rm GeV}^{-1}{\rm cm}^{-2}{\rm sec}^{-1}{\rm sr}^{-1},\nonumber \\
\Phi_{e^{-}}^{sec}(E) & =\frac{0.7E^{0.7}}{1+110E^{1.5}+600E^{2.9}+580E^{4.2}}\quad{\rm GeV}^{-1}{\rm cm}^{-2}{\rm sec}^{-1}{\rm sr}^{-1},\nonumber \\
\Phi_{e^{+}}^{sec} & (E)=\frac{4.5E^{0.7}}{1+650E^{2.3}+1500E^{4.2}}\quad{\rm GeV}^{-1}{\rm cm}^{-2}{\rm sec}^{-1}{\rm sr}^{-1},\,\label{eq:bg_e}\end{align}
where $E$ is in units of GeV. Therefore, the fraction of $e^{+}$
flux is 
\begin{equation}
\frac{\Phi_{e+}^{prim}+\Phi_{e+}^{sec}}{\Phi_{e+}^{prim}+\Phi_{e+}^{sec}+k\Phi_{e^{-}}^{prim}+\Phi_{e^{-}}^{sec}},\,\label{eq:fract_e}\end{equation}
where $k$ is a free parameter which is used to fit the data when
no primary source of $e^{+}$ flux exists\ \cite{Baltz:1998xv,Baltz:2001ir}.
Note also that the primary flux of $e^{-},$ $\Phi_{e^{-}}^{prim}$,
in the denominator of Eq.\ (\ref{eq:fract_e}) should include the
contributions from dark matter $A'$ decay as well. Our numerical
results for $m_{A'}=100$ GeV (the magenta line) and $300$ GeV (the
green line) are shown in Fig.\ \ref{fig:Fraction_e}. The prediction
of our model is consistent with the observed excess quite well, and
the position fraction starts increasing around $E\simeq20$ GeV and
$E\simeq10$ GeV for $m_{A'}=100$ GeV and $m_{A'}=300$ GeV, respectively.
Another key feature of the signal prediction is that the curve drops
off sharply at \textbf{$E=m_{A'}/2$} mainly due to the contribution
of $e^{+}$ from \textbf{$A'\to e^{+}e^{-}$} decay channel, i.e.
the peak seen in Fig.\textbf{\ \ref{fig:dnde}}. These characteristics
can be checked by the upcoming PAMELA data.

\section{Discussion and Conclusions}

\label{sec:Conclusion} 
Let us here discuss briefly how such a small kinetic mixing in Eq.~(\ref{eq:kinetic})
can arise. If there is an unbroken $Z_{2}$ parity symmetry under
which the hidden gauge boson flips its sign, the kinetic mixing would
be forbidden. This parity may arise from a more fundamental symmetry
or it may be an accidental one. In order to have a non-vanishing kinetic
mixing, we need to break the parity by a small amount and transmit
the breaking to the kinetic mixing~%
\footnote{ With the presence of the parity symmetry, a small breaking is natural
in the sense of `t Hooft. %
}. To this end we introduce messenger fields $\xi_{1}$ and $\xi_{2}$
that have both $U(1)_{Y}$ and $U(1)^{\prime}$ charges as $(1,1)$
and $(1,-1)$, respectively. Under the parity, they transform as:
\begin{equation}
\xi_{1}\leftrightarrow\xi_{2}.\,\label{eq:parity_messenger}\end{equation}
 Note that the charge assignment is consistent with the parity transformation
(see the discussion below Eq.\ (\ref{eq:gauge_scalar})). We assume
that the messenger mass scale $M_{m}$ is very large, e.g. the grand
unified theory (GUT) scale. Suppose that the parity is broken in the
messenger sector in such a way that $\xi_{1}$ and $\xi_{2}$ obtain
different masses. Let us denote the mass difference by $\Delta M_{m}$,
which parametrizes the amount of the parity violation. Integrating
out the messengers, we are then left with a small kinetic mixing,
$\epsilon$, given by 
\begin{equation}
\epsilon\;\sim\;\frac{g_{h}g^{\prime}}{16\pi^{2}}\frac{\Delta M_{m}^{2}}{M_{m}^{2}},\end{equation}
where $g_{h}$ is the coupling constant of the hidden $U(1)$ gauge
symmetry. For $\Delta M_{m}\sim{\cal O}({\rm TeV})$ and $M_{m}\sim10^{15}{\rm \, GeV}$,
we obtain the tiny kinetic mixing of the right magnitude of ${\cal O}(10^{-26})$.

It is tempting to identify the origin of the parity violation with
the spontaneous breaking of the $U(1)^{\prime}$. As an illustration,
we will present a toy model below. As we will see later, there are
some dangerous couplings which may spoil the stability of the $\ap$
in this model. However, these problems could be solved by embedding
the model into a theory with supersymmetry or an extra dimension(s).

Suppose that all the matter fields in the hidden sector are neutral
under the parity. Then, the gauge field $B_{\mu}^{\prime}$ cannot
have any interactions with those hidden matter fields, since they
are forbidden by the parity. Let us introduce two scalar fields, $\phi_{1}$
and $\phi_{2}$, which transform under the parity as \begin{equation}
\phi_{1}\leftrightarrow\phi_{2},\label{eq:parity_scalar}\end{equation}
 and we assume that $\phi_{1}$ and $\phi_{2}$ are neutral under
the SM gauge symmetries. Then the following interactions are allowed:\begin{equation}
{\cal L}\supset ig_{h}B_{\mu}^{\prime}(\phi_{1}^{*}\partial^{\mu}\phi_{1}-h.c.)-ig_{h}B_{\mu}^{\prime}(\phi_{2}^{*}\partial^{\mu}\phi_{2}-h.c.).\,\label{eq:gauge_scalar}\end{equation}
The parity is spontaneously broken if one of the two scalars develops
a non-vanishing vev. To this end, we consider the following potential:\begin{equation}
\lambda(|\phi_{1}|^{2}-v_{h}^{2})^{2}+\lambda(|\phi_{2}|^{2}-v_{h}^{2})^{2}+2\kappa|\phi_{1}|^{2}|\phi_{2}|^{2},\,\label{eq:potential}\end{equation}
 where $v_{h}$ , $\lambda$ and $\kappa$ are real and positive.
For $\kappa>\lambda$, there are four distinct vacua, $(\phi_{1},\phi_{2})=(\pm v_{h},0)$
and $(0,\pm v_{h})$. We take one possibility of them, $(\phi_{1},\phi_{2})=(v_{h},0)$,
as an example. Therefore, the hidden $U(1)$ gauge symmetry is spontaneously
broken by the vev of $\phi_{1}$, and the associated gauge boson acquires
a mass, $m=\sqrt{2}g_{h}v_{h}$, by eating the imaginary component
of $\phi_{1}$.

In order to transmit the parity violation, we introduce couplings
between $\phi_{1,2}$ and the messenger fields, 
\begin{equation}
-{\cal L}\supset M_{m}^{2}(|\xi_{1}|^{2}+|\xi_{2}|^{2})+\kappa^{\prime}(|\phi_{1}|^{2}|\xi_{1}|^{2}+|\phi_{2}|^{2}|\xi_{2}|^{2}),\,\label{eq:Scalar_messengers}\end{equation}
 where $M_{m}$ is the messenger mass, and $\kappa^{\prime}$ is a
real and positive constant. After the $U(1)^{\prime}$ is spontaneously
broken, masses of $\xi_{1}$ and $\xi_{2}$ are slightly different:
$m_{\xi_{1}}^{2}=M_{m}^{2}+\kappa^{\prime}v_{h}^{2}$ and $m_{\xi_{2}}^{2}=M_{m}^{2}$.
After integrating out these heavy messengers, we obtain the kinetic
mixing in Eq. (\ref{eq:kinetic}) with $\epsilon$ given by\[
\epsilon\sim\frac{g_{h}g^{\prime}}{16\pi^{2}}\frac{\kappa^{\prime}v_{h}^{2}}{M_{m}^{2}}.\]
 For $g_{h}\sim\kappa^{\prime}\sim{\cal O}(1)$, $v_{h}\sim1{\rm \, TeV}$,
and $M_{m}\sim10^{15}{\rm \, GeV}$, we obtain $\epsilon\sim10^{-26}$.
We can therefore realize more or less the correct magnitude of $\epsilon$
needed to account for the excesses of the gamma rays and the positrons
in this toy model.

In the above toy model, we have assumed that the SM particles couple
to the hidden sector only through the messenger fields. It is also
possible to introduce direct interactions between $\phi_{1,2}$ and
the visible sector. For instance, we can couple them to the SM fermions
as \begin{equation}
{\cal L}\supset\frac{\alpha}{M_{p}^{2}}(|\phi_{1}|^{2}+|\phi_{2}|^{2})\bar{f}_{L}f_{R}H+{\rm h.c.}\,,\label{eq:scalar_SM}\end{equation}
where $\alpha\sim{\cal O}(1)$, $M_{p}\simeq2.4\times10^{18}{\rm \, GeV}$
is the reduced Planck scale, and $f_{R(L)}$ is the right-(left-)handed
fermion and $H$ is the Higgs field. Although this interaction does
not lead to the gauge kinetic mixing, it induces the decay of the
hidden gauge boson into a fermion pair after breaking the parity.
However, the decay branching ratio through such interaction is negligible,
compared to that through the kinetic mixing with $\epsilon\sim10^{-26}$.

More dangerous direct couplings are those between $\phi_{1,2}$ and
the Higgs field, $(|\phi_{1}|^{2}+|\phi_{2}|^{2})|H|^{2}$. If there
exist such interactions, the $\ap$ could decay into the Higgs bosons
immediately. If we extend the model into a supersymmetric one, the
direct couplings to the Higgs bosons can be suppressed with the aid
of the $R$-symmetry. In a theory with an extra dimension, it is also
possible to suppress the direct couplings with appropriate configurations
of the branes (e.g. the visible particles on one brane, while the
hidden particles on the other).

It is also known that many $U(1)$ symmetries appear in the string
theory, and the kinetic mixing can similarly arise in the low energy
theory by integrating out heavy string states that have charges of
two $U(1)$ gauge symmetries. It has been extensively studied how
large the kinetic mixing can be in e.g. Ref.\ \cite{Dienes:1996zr}
(see also Ref.~\cite{Abel:2008ai} and references therein). For instance,
in a warped background geometry, we can have an exponentially small
kinetic mixing~\cite{Abel:2008ai}.

Let us also have some comments on the the GUT. So far we have assumed
the existence of the kinetic mixing between $U(1)_{Y}$ and the hidden
$U(1)^{\prime}$. If one of the $U(1)$ gauge symmetries actually
sits within an unbroken non-Abelian gauge symmetry, such kinetic mixing
is not allowed. This does not necessarily mean that kinetic mixings
are incompatible with GUT, because the GUT gauge group is spontaneously
broken. Indeed, kinetic mixings between the $U(1)^{\prime}$ and the
GUT gauge fields can arise below the GUT scale by picking up non-vanishing
vevs of the Higgs bosons responsible for the GUT breaking.

Let us also discuss how the hidden gauge boson could be generated
in the early universe to account for the observed abundance of dark
matter. For simplicity, we neglect a numerical coefficient of order
unity in the following discussion. In the presence of the messenger
fields, there appears at one-loop level a following interaction between
the $U(1)^{\prime}$ and $U(1)_{Y}$ gauge fields: 
\begin{equation}
{\cal L}\;\supset\;\frac{g_{h}^{2}g^{\prime2}}{16\pi^{2}M_{m}^{4}}(B_{\alpha\beta}B^{\alpha\beta})(B_{\gamma\delta}^{\prime}B^{\prime\gamma\delta}).\end{equation}
The hidden gauge boson $A^{\prime}$ will be produced through the
above interaction most efficiently at the reheating. The $\ap$ abundance
is roughly estimated to be 
\begin{equation}
\frac{n_{\ap}}{s}\;\sim\;10^{-12}\, g_{h}^{4}\lrfp{T_{R}}{3\cdot10^{13}{\rm \, GeV}}{7}\lrfp{M_{m}}{10^{15}{\rm \, GeV}}{-8},\end{equation}
where $s$ and $T_{R}$ are the entropy density and the reheating
temperature, respectively. For the mass $m_{\ap}\sim{\cal O}(100){\rm \, GeV}$,
a right abundance of $\ap$ can be generated from the above interaction
for $T_{R}\sim10^{13}$ GeV. Also the $\ap$ can be non-thermally
produced by the inflaton decay~\cite{Endo:2006qk,Endo:2007ih,Endo:2007sz}.

In the above model, the parity is spontaneously broken by the vev
of $\phi_{1}$. If the breaking occurs after inflation, domain walls
connecting two of the four vacua in Eq.\ (\ref{eq:potential}) will
be formed~\cite{Zeldovich:1974uw,Kibble:1976sj,Vilenkin:1981zs},
which can be the cosmological disaster. There are several means to
get around this problem, and one of which is to introduce a tiny explicit
breaking of the parity symmetry. Therefore the domain walls are not
stable, and eventually annihilate after collisions~\cite{Vilenkin:1981zs,Coulson:1995nv,Larsson:1996sp}.
Another solution is to assume that the breaking occurs before inflation.
The last one is to assume that the initial positions of the scalars
$\phi_{1}$ and $\phi_{2}$ are deviated from the origin. In the last
case, the domain walls, if formed, will be annihilated eventually~%
\footnote{ The significant amount of the gravitational waves may be produced
in the collisions of domain walls~\cite{Takahashi:2008mu}. The hidden
gauge bosons are also produced by the annihilation processes of the
domain walls.%
}.

In this paper we have considered a possibility that a hidden gauge
boson $A'$, which constitutes the dominant component of dark matter,
decays into the SM particles through the kinetic mixing term that
breaks the $Z_{2}$ parity symmetry. As a result, the branching ratios
are solely determined by the mass of the hidden gauge boson. Continuum
spectra of photons and positrons are generated from $A'\to\ell^{+}\ell^{-}$
($\ell=e,\,\mu,\,\tau$), and from the decays of hadrons, mainly $\pi^{0,\,\pm}$,
produced in the subsequent QCD hadronization process~%
\footnote{There also exist antiprotons produced from the $A'$ decay, whose
flux can also be calculated and compared with the present data. However,
we did not pursue the detailed comparison in this paper due to the
large experimental uncertainties and our poor understanding of diffusion
models.%
}. If the mass of $A'$ is about ${\cal O}(100)$ GeV and its lifetime
is of order ${\cal O}(10^{26})$ seconds, those gamma rays and positrons
from $A'$ decay may account for the observed excesses in the extra
galactic diffuse gamma ray flux and the positron fraction. Interestingly,
in our model, the spectra of the gamma rays and the positrons have
distinctive features: the absence of line emission of the gamma ray
and a sharp peak in the positron fraction. Such features may be observed
by the FGST and PAMELA satellites.

\begin{acknowledgments}
CRC and FT would like to thank M. Nojiri for useful discussions. This
work was supported by World Premier International Research Center
Initiative (WPI Initiative), MEXT, Japan. 
\end{acknowledgments}

{\it Note added:}
Very recently the PAMELA group reported a steep rise in
the positron fraction~\cite{Adriani:2008zr}, which is nicely explained by
our model.

\bibliographystyle{apsrev} \bibliographystyle{apsrev} \bibliographystyle{apsrev}
\bibliography{reference}

\end{document}